\documentclass[prd,aps,floatfix,nofootinbib,twocolumn,tightenlines]{revtex4}
\usepackage{dcolumn}% Align table columns on decimal point
\usepackage{latexsym}
\usepackage{graphicx}
\usepackage{bm}
\usepackage{epstopdf}

\def\d{\delta}
\def\f{\phi}                    %       \varphi

\def\h{\eta}

\def\j{\psi}

\def\m{\mu}

\def\o{\omega}
\def\p{\pi}                     % Also, \varpi
\def\cd{{\cal D}}
\def\ce{{\cal E}}

\def\ce{{\cal E}}
\def\Re{{\rm Re\,}}
\def\Im{{\rm Im\,}}

\long \def \blockcomment #1\endcomment{}
\def\det{{\rm det}}

\begin{document}
%%%%%%%%%%%%%%%%%%%%%%%%%%%%%%%%%%%%%%%%%%%%%%%%%%%%%%%%%%%%%%%%%%%%%%
\title{Breakdown of staggered fermions at nonzero chemical potential
\vskip-7ex\rightline{\rm SINP/TNP/06-01}\vskip 6ex
}
\author{Maarten Golterman}%
%\email{maarten@stars.sfsu.edu}
 \affiliation{Theory Group, Saha Institute of Nuclear Physics,
1/AF, Salt Lake, Kolkata 700064, India}
 \altaffiliation[Permanent address: ]{Department of Physics and Astronomy,
San Francisco State University, San Francisco, CA 94132, USA}

\author{Yigal Shamir}
%\email{shamir@post.tau.ac.il}
\author{Benjamin Svetitsky}
%\email{bqs@julian.tau.ac.il}
 \affiliation{School of Physics and Astronomy, Raymond and Beverly
Sackler Faculty of Exact Sciences, Tel~Aviv University, 69978
Tel~Aviv, Israel}

\preprint{SINP/TNP/06-01}

\begin{abstract}
The staggered fermion determinant is complex when the quark chemical
potential $\mu$ is nonzero. Its fourth root, used in simulations with
dynamical fermions, will have phase ambiguities that become acute when
$\Re\mu$ is sufficiently large. We show how to resolve these ambiguities,
but our prescription only works very close to the continuum limit. We
argue that this regime is far from current capabilities. Other
procedures require being even closer to the continuum limit, or fail
altogether, because of unphysical discontinuities in the measure. At
zero temperature the breakdown is expected when $\Re\mu\agt m_\pi/2$,
where $m_\pi$ is the pion mass. Estimates of the location of the
breakdown at nonzero temperature are less certain. 
\end{abstract}

\maketitle

The direct study of quantum chromodynamics (QCD) at nonzero chemical
potential $\mu$ faces the considerable hurdle presented by the complex
measure in the euclidean path integral
(for a review, see Ref.~\cite{review}).  This requires the
adoption of various stratagems in order to make Monte Carlo analysis
possible.  These stratagems include
Taylor expansions around $\mu=0$ \cite{Taylor};
analytic continuation from the imaginary
$\mu$ axis (where the measure is real and positive) \cite{cont}; and
multiparameter reweighting using real measures that push the complex
phases into the observables \cite{Barbour,FK1,Allton,FK2,FK3}.  All these
methods envisage a well-defined path integral with complex measure at
real, nonzero values of $\mu$.
While the first two methods never actually calculate fermion
determinants for $\Re\mu\not=0$, the third method does
calculate ratios of these determinants as configurations are analyzed.
For a different reweighting method based on the canonical ensemble,
see Ref.~\cite{FoKr}.

Calculations at $\mu\not=0$ are often performed with staggered
fermions.  These are contained in a single-component fermion field that
makes for efficient calculation of the fermion determinant
and Green functions;
a further advantage is a chiral symmetry that is only broken softly
by the mass parameter $m$.  A disadvantage of
staggered fermions is residual doubling, wherein a single lattice field
actually represents four physical flavors (nowadays called tastes) that become
degenerate in the continuum limit.  All these tastes affect the dynamics
of the theory, as is easily seen by considering fermion loops in a
diagrammatic expansion of the determinant.
In order to represent a single flavor of fermion with mass $m$,
then, one takes the fourth root of the determinant.
Thus one hopes to arrive at two light flavors
and one strange quark in the continuum limit.

The QCD partition function on a lattice, with two light quarks plus
the strange quark, is given by
\begin{eqnarray}
Z&=&\int \cd U \,e^{-S_g(U)}\,\det[D(U)+m_l]^{1/2}\nonumber\\
&&\qquad\times\det[D(U)+m_{s}]^{1/4},
\end{eqnarray}
where $D(U)$ is the staggered-fermion hopping matrix, including factors
of $\exp(\pm\mu)$ on the timelike links; $m_l$ is the common mass
of the two light quarks and $m_s$ is the strange quark mass.
When $\Re\mu=0$, the matrix $D(U)$ is anti-hermitian.
The eigenvalues of $D+m$ then come in conjugate pairs, $\eta = m\pm iE$, so
the determinant is real and strictly positive (for $m\ne 0$).
If we take the fourth root to be
real and positive as well, the measure in $Z$ is a continuous functional
of the gauge field $U$.
We will assume here that taking the fourth root is a valid procedure
for $\mu=0$ \cite{AB,rg,4fold}.

When $\Re\mu\not=0$, on the other hand, the spectrum of $D+m$ spreads
out into the complex plane and the determinant becomes complex.  Taking the
fourth root of the complex determinant introduces phase ambiguities.  A
prescription must be found for resolving these ambiguities.  The problem
first appears when $\mu\agt m$, when rare configurations admit eigenvalues near zero.
At some larger value of $\mu$, near-zero eigenvalues
become common in typical configurations.
Analysis of the $\epsilon$-regime suggests that in the zero-temperature theory this happens rather abruptly when
$\mu\approx m_\pi/2$ \cite{mpi};  at $T>0$ this point is believed to move towards higher $\mu$ \cite{Kim}.

We present below a sensible prescription for taking the fourth root when
$\Re\mu\not=0$.  This prescription requires that we
work near the continuum limit.  The breakdown of the method on
coarser lattices, in fact for any realistic value of the lattice spacing,
is quite general and indicative as well of the failure of
alternative rules for dealing with the fourth root.  The
discussion carries over to the \textit{square} root used
for the two light quarks.

\begin{figure}[thb]
%\vspace*{3ex}
\begin{center}
\includegraphics*[width=0.8\columnwidth]{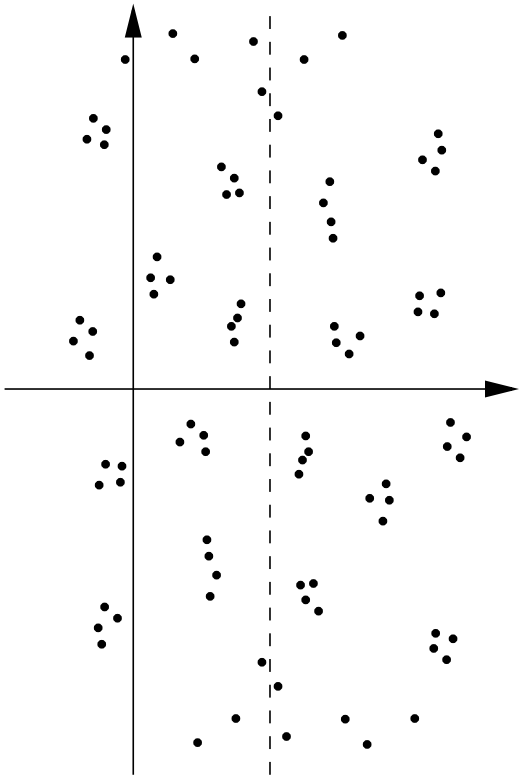}
\end{center}
\caption{Sketch of complex eigenvalues $\eta_i$ of $D+m$ when $\Re\mu$
is somewhat larger than $m_\pi/2$. The dashed line is $\Re\eta=m$.
Eigenvalues closest to the real axis are clustered into quartets
with splittings of $O(a\Lambda^2)$; typical spacing between quartets
is $\propto V^{-1/2}$
 (see text).
Farther from the real axis the quartet structure disappears.
The eigenvalues come in pairs $(\eta,2m-\eta)$ \cite{SmitVink}.
\label{quartets}}
%\vspace*{-1ex}
\end{figure}

An unambiguous definition of the phase of the fourth root requires
understanding taste symmetry and its breaking at finite lattice spacing.
To this end,
let us describe the IR and UV regimes of the complex
spectrum of $D+m$ near the continuum limit  (see Fig.~\ref{quartets}).
%%%%%%%%%%%%%%%%%%%
%%%%%%%%%%%%%%%%%%%
The eigenvalues $\eta_i$ that fall in the IR regime will form
approximately degenerate, well distinguished quartets, as dictated by
taste symmetry.  These will include eigenvalues near the QCD scale,
$|\eta_i|\approx \Lambda$.  For UV eigenvalues, with $|\eta_i|\to a^{-1}$,
large taste violations are unavoidable.  Since, however, both $m$ and
$\mu$ are now small compared to $|\eta_i|$, these eigenvalues are
dominated by the anti-hermitian $D(\mu=0)$, and so their phases should
be near $\pm\pi/2$, with deviations that are $O(am,a\mu)$.  We expect
that there will be a large overlap region where the eigenvalues $\eta_i$
satisfy both conditions: they should fall into approximately degenerate
quartets \textit{and} their phases will be near $\pm\pi/2$.

Now the prescription for the fourth root of the determinant is
straightforward.  For the IR quartets, we proceed one quartet at a time,
and we take the fourth root of their product
to have a phase angle equal to the phase of the center of the quartet.
Thus we construct a single-flavor
theory, with one eigenvalue in place of each quartet.  For the UV
eigenvalues,%
\footnote{Renormalization-group analysis \cite{rg} suggests that the log
of the UV part of the determinant contributes only a local correction to
the effective action of the gauge field; it can even be discarded,
if one compensates by renormalizing the gauge coupling.}
there is no need for subtlety; since the eigenvalues lie near the imaginary
axis, we can take the arguments of their fourth roots to lie near $\pm\pi/8$.

The fourth root so constructed will usually be a well-defined, continuous
functional of the gauge field, with a slowly varying phase angle.  An
exception will arise if an eigenvalue quartet comes too close to the
origin.  If, in fact, the four eigenvalues frame the origin, the average
phase of the quartet becomes meaningless and any definition will
fluctuate rapidly as the eigenvalues move.
Such an occurrence will, however, be rare.  The typical splitting $\delta$
within a quartet will be%
\footnote{This can be seen from the representation of the staggered operator
in taste basis, where the taste breaking comes from a dimension-5 operator
that shifts eigenvalues within a quartet as
$\eta\to\eta\pm\delta$, where $\delta$ is fixed and $O(a)$.
This is consistent with the usual picture of $O(a^2)$ discretization errors
because the product of eigenvalues will shift by $O(\delta^2)$.}
$O(a\Lambda^2)$, and so
the probability of finding an eigenvalue this close to zero is proportional
to $a^2$.  A more detailed estimate follows.

First, let us discuss orders of limits.
At $\m=0$,
the relative taste splittings inside a quartet,
$|\eta_i-\eta_j|/|\overline\eta|$, are bounded from above by $a\Lambda^2/m$.
If we want the taste splittings of the pseudo-Goldstone bosons to be small,
we must take the continuum limit before the chiral ($m\to0$) limit \cite{CB}.
Let us now assume $\Re\mu\sim\Lambda$. The eigenvalue density
near the origin in the complex $\eta$ plane scales like $V\Lambda^2$
with the volume $V$ of the lattice.
Thus $m$ gets replaced by the $O[(\sqrt{V}\Lambda)^{-1}]$
mean distance between quartets (the same estimate applies to the typical
magnitude of the eigenvalue nearest to the origin),
and $|\eta_i-\eta_j|/|\overline\eta|$ comes out
to be $O(a\sqrt{V}\Lambda^3)$ for the near-zero quartets.
Now it is the product $a\sqrt{V}\Lambda^3$ that must remain small for the sake
of good taste.
If not, the splittings within a quartet will
be as large as the splittings between different quartets,
and our prescription will break down.
When approaching the continuum limit,
the volume must be limited such that
$a\sqrt{V}\Lambda^3\to 0$ as well.

With the eigenvalue density estimated as above,
the fraction of \textit{quenched} configurations containing an
eigenvalue quartet that frames the origin is $O(a^2V\Lambda^6)$.
This, then, is the order of the error introduced
by either taking or dropping these configurations.
Reweighting by the determinant provides further suppression
by $O(a\sqrt{V}\Lambda^3)$;  this is the ratio of fourth roots
of the said eigenvalue quartet and of a typical near-zero quartet.
The systematic error thus becomes $O[(a\sqrt{V}\Lambda^3)^3]$.

Is it really necessary to extract individual eigenvalues and to group
them into quartets?  The procedure appears cumbersome, but other
prescriptions that come to mind produce larger errors
close to the continuum limit, or even fail altogether.
If one were to calculate the determinant $\cd$ in full and only then to take
its fourth root $\Delta$ in the cut plane $|{\rm Arg}\,\cd|<\pi$,
the result would be restricted to the wedge $|{\rm Arg}\,\Delta|<\pi/4$.
This is inconsistent with an ordinary single-flavor theory,
whose determinant is free to wander the entire complex plane.

If one were to take the fourth root of each individual $\eta_i$
in the cut plane before multiplying them together, there would be frequent
jumps of $\pi/2$ in the phase of $\Delta$ as single eigenvalues cross the
cut.  Here the systematic error can be estimated: It would be
$O(aV\Lambda^5)$, which is the fraction
of configurations in which a quartet is bisected by the negative real
axis, close to the continuum limit.
[The probability that none of the $O(V\Lambda^4)$ IR eigenvalues fall
in a strip of width $a\Lambda^2$ along the negative real axis scales like
$(1-a\Lambda)^{V\Lambda^4} \approx 1-aV\Lambda^5$.]
This estimate is only valid if one has verified the existence of well-defined quartets in all configurations.
Moreover, this error is can be fatally large even if the figure of merit $a\sqrt{V}\Lambda^3$ is small.

Are current calculations
anywhere near to realizing the quartet structure of IR eigenvalues
that will make calculation of the fourth root consistent?
We suspect that the answer is negative.
Simulations at $\mu=0$ show
nice taste degeneracy in the lowest eigenvalues only when
the lattice cutoff is $a^{-1} \approx 2$~GeV or larger,
and only if highly improved gauge and staggered actions are used \cite{4fold}.
The state of affairs is much less satisfactory in
existing simulations with $\Re\mu>0$, where
the cutoff is at 1~GeV or less.
Even in the $\mu=0$ simulations,
significant taste violations will reappear if one looks beyond the lowest
few quartets. While this does not hamper simulations at $\mu=0$,
for the $\mu>0$ case we do require a distinct quartet structure all the way
up to $|\eta|\sim\Lambda$ and beyond.

In general terms, what would it take to define a fourth root that is
single-valued and continuous?  Consider a closed trajectory in
gauge-field space that avoids the zeros of the fermion determinant
$\cd$.  When this trajectory is mapped into the complex $\cd$ plane, the
winding number around the origin must be a multiple of 4.  In practice,
this will never be guaranteed, precisely because of taste-symmetry
breaking.  We believe that current simulations are plagued by
eigenvalues with no particular multiplet structure.  Problems will arise
when any single eigenvalue circles the origin.  Depending on the
treatment given the fourth root, either the eigenvalue will cross a cut,
causing a jump in the fourth root; or the eigenvalue will move onto
another Riemann sheet, giving a multi-valued root.

[We note that, for $m=0$, \textit{two} eigenvalues will always circle
the origin simultaneously, because they appear in $(\eta, -\eta)$ pairs
in that case (see Fig.~\ref{quartets}).  It would then be possible to
define a \textit{square} root by requiring continuity, because the
winding number of the determinant around the origin would always be
even.  This only works, however, for $m=0$, and it will not help for the
fourth root.]

We illustrate this problem in two dimensions.
We shall exhibit eigenvalues near the origin for selected gauge fields
and show how small changes in the gauge field can take a single eigenvalue
around the origin and thus through any cut that can be defined.
The staggered fermion action on a lattice of $L_t\times L_x$ sites is
\begin{eqnarray}
  S &=&\bar\j(D+m)\j\nonumber\\
&=&  \frac12 \sum_{t=1}^{L_t}\sum_{x=1}^{L_x} \left[
  \bar\j_{t,x}\, U^0_{t,x}\, \j_{t+1,x}
  -\bar\j_{t+1,x}\, U_{t,x}^{0\dagger}\, \j_{t,x}\right.
\nonumber\\
  &&\qquad+(-1)^t
  \left(\bar\j_{t,x}\, U^1_{t,x}\, \j_{t,x+1}
  -\bar\j_{t,x+1}\, U_{t,x}^{1\dagger}\, \j_{t,x}\right)
\nonumber\\
  &&\qquad  +2m\bar\j_{t,x}\j_{t,x}\Bigr].
\label{2d}
\end{eqnarray}
A change of variables \cite{RW}
has removed the chemical potential from all links
of the lattice except for those that connect $t=L_t$ to $t=1$.
First we choose a gauge field that is zero except for
a constant Polyakov loop $U$
on the last time-link.  This is equivalent to writing the free operator
with the modified antiperiodic boundary conditions
\begin{equation}\begin{array}{ll}
  \j_{t,L_x+1} \equiv  -\j_{t,1}\,,&
  \qquad  \j_{L_t+1,x} \equiv  - U e^{L_t\m}\j_{1,x}\,,\\[3pt]
  \bar\j_{t,L_x+1} \equiv - \bar\j_{t,1}\,,&
  \qquad \bar\j_{L_t+1,x} \equiv - \bar\j_{1,x} U^\dagger e^{-L_t\m}\,.
  \end{array}
\label{BCf}
\end{equation}
\mbox{}\vskip0.5ex\noindent
A Fourier transformation from $x$ to $p$ gives
\begin{eqnarray}
  S &=&
  \sum_{t=1}^{L_t}\sum_{p} \left\{
  \frac12 \left(\bar\j_{t,p}\, \j_{t+1,p} -\bar\j_{t+1,p}\, \j_{t,p}\right)
\right.\nonumber\\
  &&\qquad+ \left[m+(-1)^t i\hat p\right] \bar\j_{t,p}\j_{t,p}\bigg\}
\nonumber\\
  &\equiv& \sum_{t,t'}\sum_p \bar\psi_{t,p} M_{t,t'}(p) \psi_{t',p},
\label{2dp}
\end{eqnarray}
where $\hat p = \sin p$.
The momentum takes the values $p_n = (2n+1)\p/L_x$.  We must keep careful
track of multiplicities. If we take $L_x$ to be a multiple of four,
then none of the momenta
$p=0,\p/2,\p,-\p/2$ are in the spectrum. This assures that each
possible value of $\sin^2p$ has a multiplicity of four.
The fermion matrix in the momentum basis is given explicitly by
%\newpage
\begin{widetext}
\begin{displaymath}
  M(p) =
  \left(\begin{array}{ccccccc}
m-i\hat p & \frac{1}{2} & 0 & 0 & \cdots & 0 & \frac{1}{2} e^{i\f-L_t\m}\\
-\frac{1}{2} & m+i\hat p & \frac{1}{2} & 0 & \cdots & 0 & 0 \\
0 & -\frac{1}{2} & m-i\hat p & \frac{1}{2} & \cdots & 0 & 0 \\
0 & 0 & -\frac{1}{2} & m+i\hat p & \cdots & 0 & 0 \\
\vdots & \vdots & \vdots &\vdots & \ddots & \vdots & \vdots \\
0 & 0 & 0 & 0 & \cdots & m-i\hat p & \frac{1}{2}\\
-\frac{1}{2} e^{-i\f+L_t\m} & 0 & 0 & 0 & \cdots & -\frac{1}{2} & m+i\hat p
  \end{array}\right) .
%\label{M}
\end{displaymath}
%\newpage
\end{widetext}
%
%\mbox{}
%\clearpage
%\pagebreak
\noindent
We have chosen $U$  diagonal,
whereupon the determinant factors into $N_c$ color factors,
and selected one of the colors.

$M$ is not anti-hermitian; it satisfies instead
$M^\dagger(m,\m) = -M(-m,-\m^*)$.
The right-eigenvectors of $M$ may be found via an alternating ansatz,
\begin{equation}
  \j_t = \left\{ \begin{array}{l}
    \alpha e^{i\o t}\,,  \qquad t \mbox{ odd,} \\
    \beta e^{i\o t}\,,  \qquad t \mbox{ even}.
  \end{array}\right.
\label{R}
\end{equation}
We find the eigenvalues
\begin{equation}
  \h_\pm = m \pm i\ce ,
\label{etapm}
\end{equation}
where $\ce = \sqrt{\hat p^2+\hat\omega^2}$,
with $\hat\omega\equiv\sin\omega$.
The boundary conditions lead to the quantization of $\Re\o$, viz.,
\begin{equation}
  \Re\o_n = \Im\mu+{(2n+1)\p-\f \over L_t},
\label{qntzR}
\end{equation}
and further require
\begin{equation}
  \Im\o_n = -\Re\m.
 \label{qntzI}
\end{equation}
The eigenvalues depend on $p$ only via $\hat p^2= \sin^2p$,
and thus the complete set of eigenvalues shows
a four-fold degeneracy.

When $\Re\mu\agt m$ [more precisely, $\cosh(2\,\Re\mu)>1+2m^2$],
the eigenvalue $\eta_+$ can be zero if $\ce$ is pure imaginary,
i.e., $\ce=im$.  This can be accomplished by giving the Polyakov loop $U$
a value corresponding to
\begin{equation}
\phi=L_t\Im\mu+\pi\qquad ({\rm mod}\,2\pi);
\end{equation}
the zero eigenvalue is then found by choosing $p$ appropriately
and taking $n=0$ so that $\Re\omega=0$.
A small change in $U$ from this value gives us
a four-fold degenerate eigenvalue near but not at zero;
in fact, $p$ cannot be chosen to give an exact zero in the first place
unless $L_x$ is infinite.

\begin{figure}[tb]
%\vspace*{3ex}
\begin{center}
\includegraphics*[width=0.8\columnwidth]{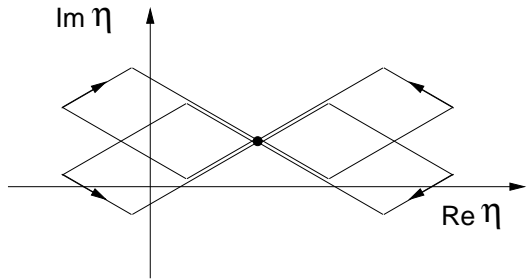}
\end{center}
\caption{Breaking of the fourfold degeneracy and transport of one eigenvalue
around the origin.
The small filled circle marks the original (fourfold) eigenvalue $\h$.
  The two upper trajectories have been
  slightly displaced for clarity.
\label{figwind}}
%\vspace*{-1ex}
\end{figure}

We can add a gauge field to break the degeneracy by pushing
the four eigenvalues in different directions.
Consider a spatial gauge field that, for all $x$,
takes values $U^1_o=e^{iA_o}$
and $U^1_e=e^{iA_e}$ for odd and even $t$ respectively.
(We assume that these spatial links commute
with the constant Polyakov loop.)
In Eq.~(\ref{2dp}), this replaces $\hat p=\sin p $ by
$\hat p_o\equiv\sin(p+A_o)$ and by $\hat p_e\equiv\sin(p+A_e)$ for odd
and even $t$ respectively. The spectrum can again be calculated exactly.
The new eigenvalues are
\begin{equation}
  \h'_\pm = m + i(\hat p_e-\hat p_o)/2 \pm i\ce' ,
\label{etapmx}
\end{equation}
where
%\begin{equation}
$\ce' = \sqrt{(\hat p_o+\hat p_e)^2/4+\hat\omega^2}$.
%\label{cex}
%\end{equation}
The conditions (\ref{qntzR}) and (\ref{qntzI}) on $\omega$ are unchanged.
If we expand to first order in $A_o$ and $A_e$, we obtain
$\h'_\pm =\h_\pm+\d\h_\pm$, where
\begin{equation}
  \d\h_\pm = {i\cos p\over 2}
  \left[ (A_e-A_o) \pm {\sin p\over \ce} (A_o+A_e) \right].
\label{detax}
\end{equation}
Equation~(\ref{detax}) contains both $\cos p$ and $\sin p$.
This means that the four-fold degeneracy among $p$, $-p$, $\p+p$
and $\p-p$ is completely lifted by the gauge field (see Fig.~\ref{figwind}).%
\footnote{When the gauge field is zero the fourfold degeneracy is a result
of taste symmetry plus charge conjugation.
The constant gauge field $A_e=A_o$ breaks the latter,
leaving the degeneracy between $p$ and $p+\pi$.  Only when $A_e\ne A_o$
is the taste symmetry broken, as can be shown with the method
of Ref.~\cite{gs}.}
%%%%%%%%%%%%%%%%%%%
%%%%%%%%%%%%%%%%%%%

To see how we can make a single eigenvalue circle the origin as in
Fig.~\ref{figwind} let us begin again from the degenerate case.
The eigenvalue quartet closest
to zero will be of order $1/L_x\ll1$.
We turn on $A_o$, then $A_e$,
and then turn them off in the same order.
The closed trajectory of only one eigenvalue circles the origin once;
hence also the determinant circles the origin once.

A {\em local\/} fluctuation in the gauge field can accomplish the same
lifting of the degeneracy and transport of eigenvalues.
For instance, beginning with the degenerate case, choose a single spacelike
link at $(t,x)$ and set $U^1_{t,x}=1+iA$, with $A$ chosen to be diagonal
in color.  The first-order change in a zero eigenvalue is
\begin{eqnarray}
  \d\h = {iA \cos p \over L_x L_t} \times
  \left\{ \begin{array}{rl}
  -1-{\displaystyle i\over \displaystyle m}\sin p , \ \ & t \mbox{ odd},\\[6pt]
    1-{\displaystyle i\over \displaystyle m}\sin p, \ \ &  t \mbox{ even}.
  \end{array}\right.
\label{deta}
\end{eqnarray}
Again, the result depends on both $\cos p$ and $\sin p$, so that
the four-fold degeneracy is lifted.

Our study of two-dimensional QCD exhibits nondegenerate
eigenvalues and shows how they wind around the origin.
The fourth root of the determinant will be problematic here; the same will apply to any representation of the determinant as
a product,
\begin{equation}
{\rm det\,}(D+m)=\prod_i \xi_i
\end{equation}
where the factors are complex functions of the gauge field, if
the fourth root of the determinant is taken by multiplying the fourth
roots of the individual $\xi_i$'s.

An example is the formula used by Fodor and Katz \cite{FK1,FK3}, where
$\xi_i=e^{L_t\mu}-\lambda_i$; the $\lambda_i$ are eigenvalues of
a non-hermitian matrix on a three-dimensional sublattice.
In our two-dimensional example, again one finds that by turning
on the gauge field $A_{e,o}$ the degeneracy of the $\lambda_i$
is completely lifted.

The method of Ref.~\cite{FK1} guarantees
continuity of the fourth root of each $\xi_i$
as a function of $\mu$, starting from a point where $\Re\mu=0$.
This is not the same as continuity as a
function of the gauge field, which, as we have seen, is the central issue.
Each $\xi_i$ carries within it a tangle of information coming from
all length scales;  we do not even know if the $\xi_i$'s will fall
into well-distinguished quartets close to the continuum limit.
If they do, it would follow as above
that the systematic error of this prescription is due
to $\xi_i$'s located a distance of order $a$ from the negative real
axis.
As noted, this gives a systematic error of $O(aV\Lambda^5)$.
If there are no quartets then the error will be larger.

We can make our estimates semi-quantitative for the simulation
parameters of Ref.~\cite{FK3}.
It is convenient to identify the QCD scale $\Lambda$ with
the critical temperature $T_c$ at $\mu=0$.
Since the temporal size of the lattices used was in the neighborhood of $1/T_c$,
in these units the lattice spacing was $a\approx1/L_t=1/4$;
the largest volume used was $(12/4)^3=27$.
As we have seen, if the eigenvalue spectrum is to exhibit a taste-quartet structure then the figure of merit $a\sqrt{V}\Lambda^3$ must be small.
For these parameters, however, we arrive at a value of 1.3.
This returns us to a picture of no quartet structure and hence an intractable error due to phase discontinuities.

The remaining question regarding Ref.~\cite{FK3} is whether the values of $\mu$ studied put the theory into the dangerous regime where the distribution of eigenvalues $\eta_i$ or $\xi_i$ reaches the origin.  In fact $\Re{\mu}$ was varied up to and beyond the putative critical end point at $\mu_c\simeq120$~MeV.%
\footnote{This is the quark chemical potential; Ref.~\cite{FK3} gives
the baryon chemical potential, $\mu_B\equiv3\mu$.}
This is well above the value of $m_\pi/2$ for the realistic light-quark masses used, but, as we have noted, the critical value of $\mu$ is probably larger when $T>0$.  The fact that the pseudo-critical temperature $T_c(\mu)$ determined in Ref.~\cite{FK3} agrees over a substantial interval with that calculated via analytic continuation \cite{cont} is indicative of such a shift, perhaps as far as $\mu\simeq100$~MeV.
Direct study of the spectral density would settle this question.

Let us summarize our findings.
While there are new complications,
taking the fourth root when $\Re\mu\not=0$ could \textit{in principle}
be on a footing similar to that of the $\m=0$ case. A requirement is
that one work close enough to the continuum limit \textit{and} set the volume
such that taste multiplets have reappeared in the
eigenvalue spectrum all the way up to $|\eta|\sim\Lambda$ and beyond.
If one tracks the quartets and assigns their phases carefully, then configurations with quartets that frame the origin will bring a systematic error of $O[(a\sqrt{V}\Lambda^3)^3]$.
If one calculates fourth roots of individual eigenvalues and ignores their quartet structure,
then quartets that are bisected by the cut cause an error of $O(aV\Lambda^5)$.
In practice, however, the demonstration of the necessary
quartet structure
appears to be far beyond any realistic simulation even at $\Re\m=0$,
let alone when $\Re\mu\not=0$.

In simulations done at purely imaginary $\m$ \cite{cont}
the validity of the fourth root is on the same
footing as at $\m=0$, provided the continuum extrapolation is done
before attempting the analytic continuation in $\mu$.
Otherwise, on a finite lattice,
the integration over the compact gauge field commutes with
the analytic continuation in $\mu$. The method would then be tantamount to
some prescription for taking the fourth root of the determinant
on the individual configurations.
The nature of this prescription, and its relation to the ones compared here,
is beyond the scope of this paper.

\begin{acknowledgments}
We thank Philippe de Forcrand for stimulating our interest in this subject
and for his comments. We also thank Zoltan Fodor and Kim Splittorff for
timely and helpful correspondence and much discussion.
We would like to thank
the organizers of the Cyprus Workshop on Computational Hadron Physics
(September 2005),  where this work was initiated.
MG thanks the physics departments at Tel Aviv University
and the University of Rome ``La Sapienza'' as well as the
Institute for Mathematical Sciences in Chennai for hospitality.
This work was supported by the Israel Science Foundation under grant
no.~173/05 and by the US Department of Energy.
\end{acknowledgments}

\end{document}